\documentclass[twocolumn,times,twocolappendix]{aastex6}
\usepackage{epstopdf,natbib}
\usepackage{graphicx}
\slugcomment{Draft version November 29, 2016}
\shorttitle{Infrared spectrum of HC\MakeLowercase{l}$^+$}
\shortauthors{Dom\'enech et al.}

\begin{document} 
\newcommand{\wn}{cm$^{-1}$}

\title{The high resolution infrared spectrum of HC\MakeLowercase{l}$^+$}

\author{J. L. Dom\'enech} \affil{Molecular Physics Department, Instituto de Estructura de la
Materia (IEM-CSIC), Serrano 123. E-28006 Madrid, Spain}
\author{B. J. Drouin} \affil{Jet Propulsion Laboratory, California Institute of Technology,
4800 Oak Grove Drive, Pasadena, CA 91109-8099, USA}
\author{J. Cernicharo} \affil{Molecular Astrophysics Group, Instituto de Ciencia de
Materiales de Madrid (ICMM-CSIC), Sor Juana In\'es de la Cruz 3, Cantoblanco, E-28049 Madrid,
Spain}
\author{V. J. Herrero, I. Tanarro} \affil{Molecular Physics
Department, Instituto de Estructura de la Materia (IEM-CSIC). Serrano 123. 28006 Madrid,
Spain} \email{jl.domenech@csic.es}

\begin{abstract}

The chloroniumyl cation, HCl$^+$, has been recently identified in space from {\it Herschel}'s spectra. A joint analysis of extensive vis-UV spectroscopy emission data together with a few high-resolution and high-accuracy millimiter-wave data provided the necessary rest frequencies to support the astronomical identification. Nevertheless, the analysis did not include any infrared (IR) vibration-rotation data. Furthermore, with the end of the {\it Herschel} mission, infrared observations from the ground may be one of the few available means to further study this ion in space. In this work, we provide a set of accurate rovibrational transition wavenumbers as well as a new and improved global fit of vis-UV, IR and millimiter-wave spectroscopy laboratory data, that will aid in future studies of this molecule. 
\end{abstract}


\section{Introduction} The study of interstellar hydrides has received a great push in the recent years, much of it due to observations made from the {\it Herschel Space Observatory}. Having relatively large rotational constants, many rotational transitions of hydrides lie in the millimiter- and sub-millimitr wave regions, which are difficult, albeit not impossible, to access from ground-based observatories, even for sites as good as those of ALMA or APEX. Since hydrides are some of the first molecules to form in space from atomic gas and molecular hydrogen, they provide invaluable information about the environment in which they are found \citep{Ger16}. Among some of the recent molecules identified in {\it Herschel}'s spectra are the chlorine-bearing compounds H$_2$Cl$^+$ \citep{Lis2010} and HCl$^+$ \citep{DeLuca2012}. The related HCl neutral had already been identified in the interstellar medium by \citet{Blake85}. H$_2$Cl$^+$ has later been detected in a variety of lines of sight \citep{Neufeld12, Neufeld15}, both in absorption and emission, also from {\it Herschel}'s spectra, and in an extragalactic source in ALMA spectra \citep{Muller14}. The only published detection of HCl$^+$ that we are aware of is that reported by \citet{DeLuca2012} in the line of sight of the luminous continuum sources W49N and W31C near the galactic center.

The key chemical pathways leading to the formation and destruction of chlorine-containing molecules in the insterstellar medium (ISM) are described in \citet{Neufeld09} and references therein. Succinctly, since the ionization potential of Cl is 12.8 eV, slightly lower that that of H, chlorine is mostly in ionized form in the diffuse medium, and it is known to react exothermically with H$_2$ to form HCl$^+$. Further reaction with H$_2$ forms H$_2$Cl$^+$, which can produce neutral HCl or Cl upon electron recombination. Chemical models \citep{Neufeld09} predict the highest concentrations of HCl$^+$ in regions with low extinction $A_v\simeq0.8$), with molecular H$_2$ accounting for less than 1 \% of the available H.

The identification of HCl$^+$ by \citet{DeLuca2012}, relied on the calculated transition frequencies and hyperfine patterns derived from a Dunham-type isotope-independent fit of two sets of laboratory data \citep{Gupta12}: the first one was an extensive set of vis-UV emission measurements \citep{She72} comprising more that 8000 transitions in the $A^2\Sigma-X^2\Pi$ band of all isotopologues. The precision of those lines was $\sim0.04$ \wn, and the spectral resolution was too low to provide any information on the hyperfine structure (hfs). The second set of data \citep{Gupta12} provided 34 accurate frequencies in the millimiter-wave region, resolving the chlorine hyperfine patterns of three rotational transitions for H$^{35}$Cl$^+$ and H$^{37}$Cl$^+$. This allowed \citet{Gupta12} to perform a joint analysis of both sets of data yielding highly accurate predictions, enough to identify the HCl$^+$ transitions in {\it Herschel}'s spectra. However, the fit did not contain any rotationally resolved data from vibrational transitions in the ground state. Adding that information to the global fit will improve the accuracy of the predictions of the model for other transitions and may facilitate possible observations in the future in the infrared (IR) region of the spectrum. Indeed, IR observations from ground platforms at high spectral resolution are a way to build up in the study of interstellar hydrides, providing an alternative and complementary tool to the millimiter and sub-millimeter observations, especially now that the {\it Herschel} mission is over. The instrument GREAT onboard SOFIA covers the frequency region of the lowest-frequency rotational transitions of H$^{35}$Cl$^+$ and H$^{37}$Cl$^+$, although telluric lines at 1441.1 and 1444.0 GHz overlap some of the hyperfine components of the transitions and might hinder the observations.  Nevertheless, there are current observation programs of HCl$^+$ with GREAT toward Sgr B2 (M), W31C (G10.6-0.4), G29.96-0.02, W49N, W51, W3(OH), and NGC6334I \citep{2015sofi.prop...20N,2016sofi.prop...16N}.  

In this paper we provide a set of accurate wavenumbers for vibration-rotation transitions of the $v=1\leftarrow0$ band of H$^{35}$Cl$^+$ and H$^{37}$Cl$^+$ in the mid-IR, as well as an extended and improved fit with millimeter-wave, optical and IR data. There is one previous laboratory IR study by diode laser spectroscopy, described in a preliminary note \citep{Dav86}, where 20 fine transitions of H$^{35}$Cl$^+$ and 13 fine transitions of H$^{37}$Cl$^+$, were measured with an uncertainty of $\pm0.002$ \wn. The limited number of observations was most likely due to the patchy coverage of lead salt diode lasers. On account of the much more extensive and accurate set of wavenumbers obtained in this investigation, those results have not been included in the present work. 

\section{Experimental details}

The apparatus used in this experiment is basically the same that has been used recently for the investigation of the IR spectra of NH$_3$D$^+$ and ArH$^+$ ions \citep{Dom13,Cue14}. It is based on an IR difference-frequency laser spectrometer, a hollow-cathode discharge reactor, and a double modulation technique with phase-sensitive detection. The only major change with respect to those previous works is the substitution of the LiNbO$_3$ crystal for a MgO-doped periodically poled LiNbO$_3$ chip with different poling periods, covering without gaps the 1900-4300 \wn region when pumped by the 514 nm line of an Ar$^+$ laser, and providing significantly more IR power than our previous setup ($\sim1$ mW vs. $\sim1 \mu$W.) The accuracy and repeatability of the frequency scale rely on the active frequency locking of the Ar$^+$ laser to an hyperfine transition of $^{127}$I$_2$, and a high-accuracy wavemeter, calibrated with the same laser, resulting in a 3$\sigma$ accuracy of 10 MHz (3.3$\times10^{-4}$ \wn ) in the IR frequency scale. The instrumental resolution is $\sim3$ MHz (10$^{-4}$ \wn), so the observed linewidths are limited by the Doppler effect.

In most of the experiments, the cathode was water-cooled, although, in a second step, some lines were recorded again using a flow of liquid nitrogen. A thermocouple resting at the mid-point of the cathode read $\sim$330 and $\sim$180 K, respectively, under discharge conditions. The plasma was operated in a flow of HCl diluted in He, passed through a dry-ice/ethanol trap to remove possible water traces. The strongest signals were obtained with a partial pressure of He of 1.2 mbar, and a partial pressure of HCl below the resolution of the available capacitance pressure meter (0.001 mbar).  The discharge current was 325 mA, with 400 V rms between electrodes. The rest of operating conditions are similar to those reorted in \citet{Cue14}.

Line positions were calculated with the polynomial expressions given by \citet{She73}, that do not account for any hyperfine splitting, and later they were taken from the prediction available at the JPL database \citep{JPLcatalog,intro_JPL-catalog}, entry c036005, with frequencies for all hyperfine components. The accuracy was high enough to find the lines within less than 0.02 \wn from their expected positions. Between 6 and 200 scans were averaged for each line, depending on its intensity, attaining signal-to-noise ratios between 10 and 500. The investigated region spanned the 2337-2774 \wn interval. Atmospheric CO$_2$ absorption hampered the detection at lower frequencies (higher-$J$ lines in the $P$-branch), and we kept a similar $J''$ range for the $R$-branch. No attempt was made to cancel the earth's magnetic field in the experiments (44.6 $\mu$T in Madrid), since the calculation of Zeeman energies by \citet{che07} showed a worst-case splitting (that of the $^2\Pi_{3/2}\ Q(3/2)$ lines) of only 3 MHz, much smaller than their Doppler width.  

\section{Spectroscopic results}
\begin{figure}[tbp!]
\begin{center}
\includegraphics[width=\columnwidth]{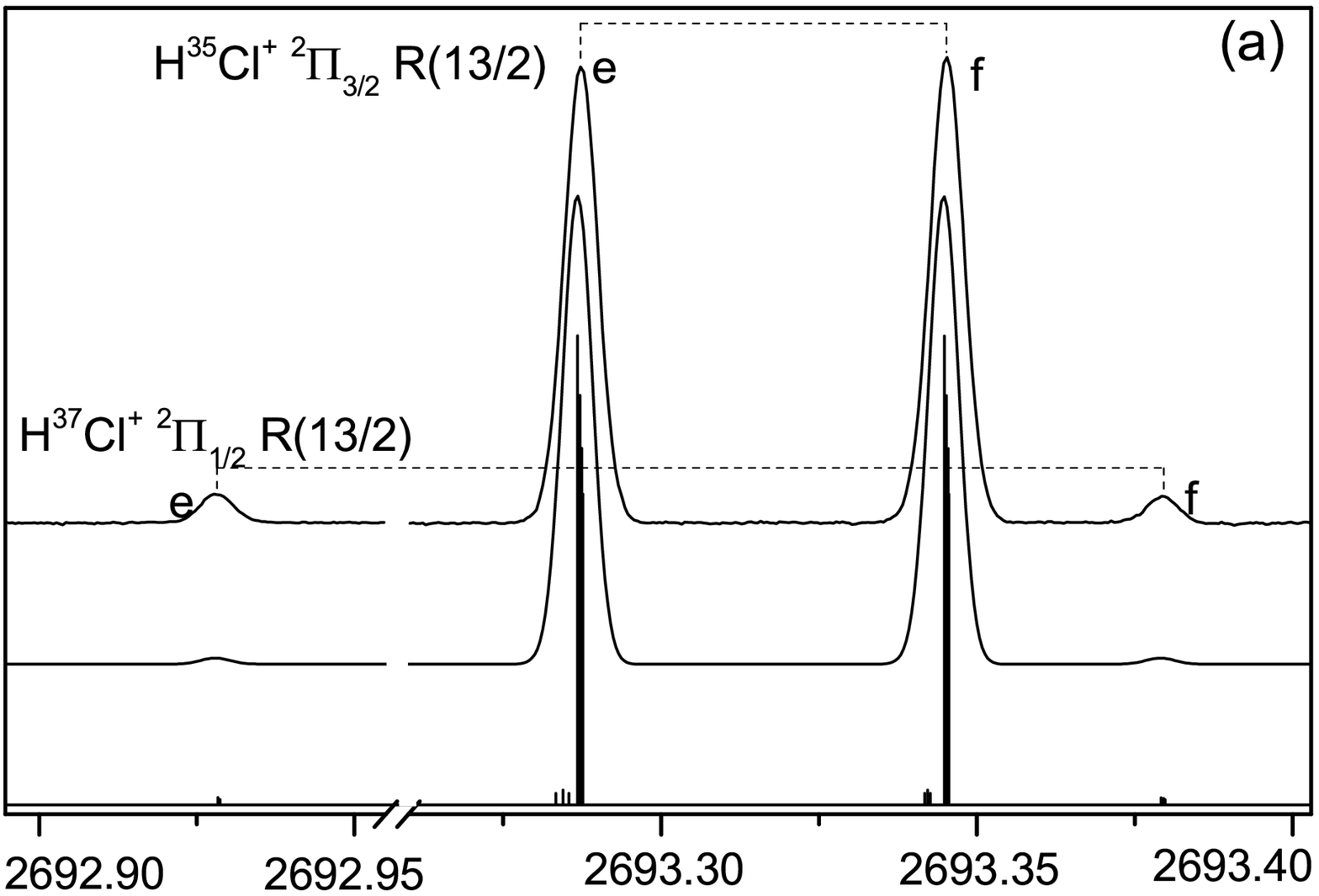}
\includegraphics[width=\columnwidth]{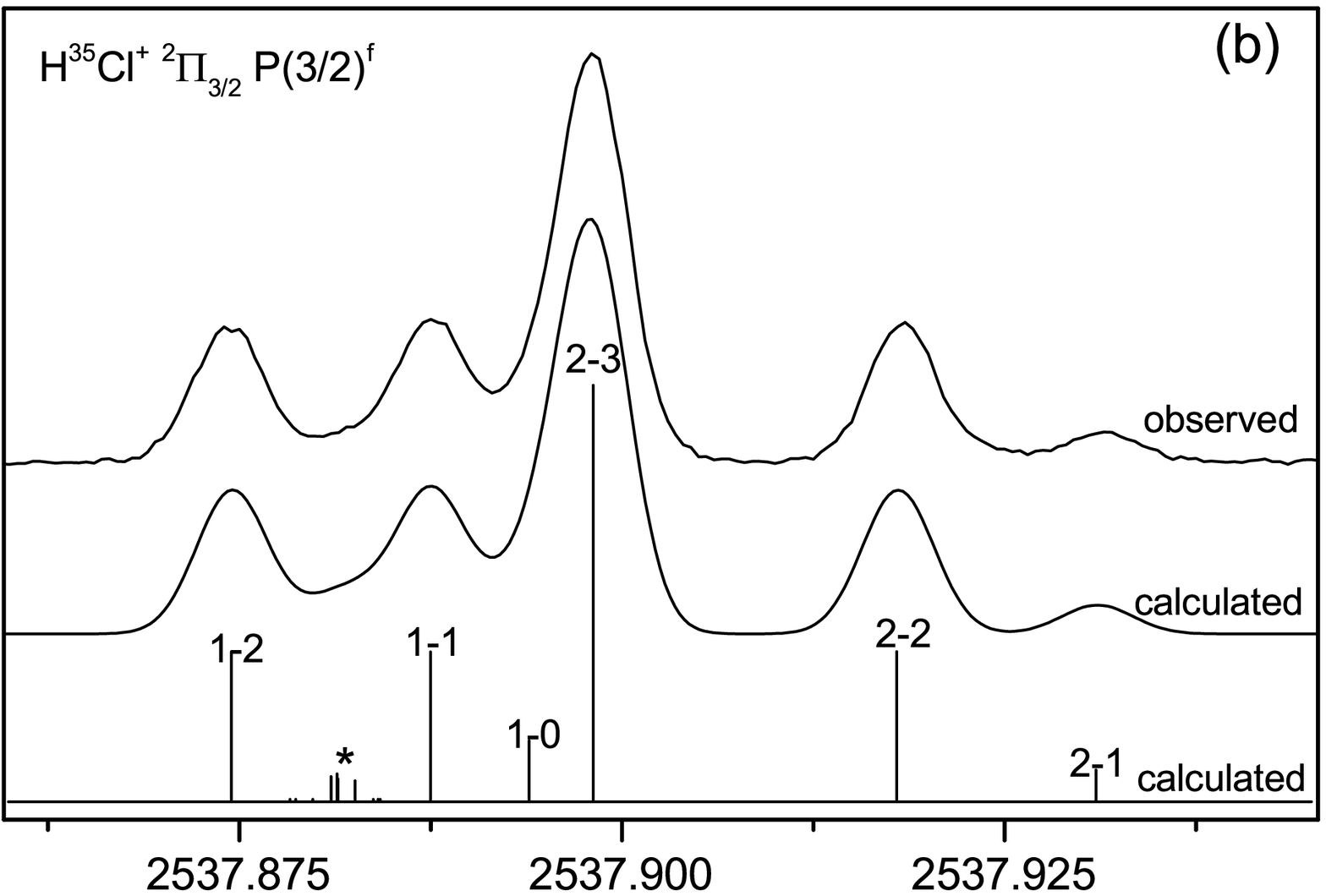}
\includegraphics[width=\columnwidth]{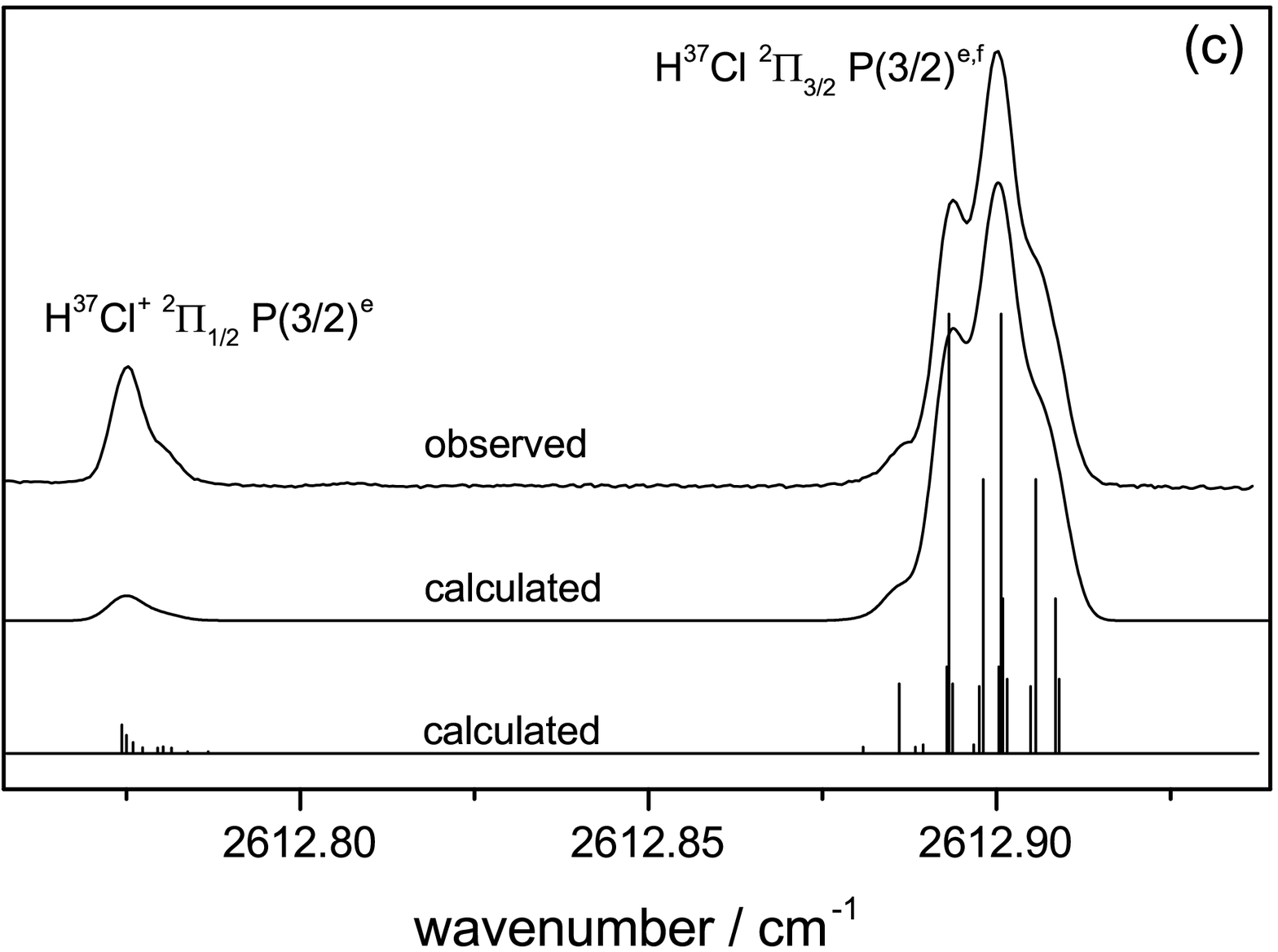}
\end{center}
\caption{Absorption lines of HCl$^+$.  In all panels, the bottom trace is a stick spectrum as derived from the fit showing the hfs, and the top trace is the experimental observation.  (a) The middle trace is the convolution of the stick spectrum with a Gaussian function of 0.0065 \wn FWHM. (b) The numbers on the top of the sticks are the total angular momentum quantum numbers $F'-F''$. The group of lines marked with an asterisk correspond to $^2\Pi_{3/2}Q(9/2)^f$.  The middle trace is the stick spectrum convolved with a Gaussian function of 0.0055 \wn FWHM. (c) The middle trace is the convolution of the stick spectrum with a Gaussian function of 0.0055 \wn FWHM. In this case, the $e$ and $f$ components in the $\Omega=3/2$ ladder are not resolved, unlike those in the $\Omega=1/2$ ladder.  }
\label{fig:figf}
\end{figure} 

\subsection{The spectrum of HCl$^+$} 

The energy levels structure of HCl$^+$ has been described in previous papers \citep[e.g.] [and references therein]{Gupta12}, so we just summarize it here. The ground electronic state of HCl$^+$ is $^2\Pi$, so there is electron spin and orbital angular momenta interaction, originating two ladders of levels with $\Omega=\left|\Lambda+\Sigma\right|...\left|\Lambda-\Sigma\right|=3/2,1/2$. Since the ratio of the constants $A/B$ is -66, it is an {\it inverted} doublet (the $\Omega=3/2$ state lies lower in energy ---by $\sim 643$ \wn) that can be well described by Hund's coupling case {\it a}. Levels in each ladder are split by $\Lambda$-doubling into $e$ and $f$ parity doublets that are further split into hyperfine components because of the interaction with the nuclear spins ($I_{^{35}{\rm Cl}}=I_{^{37}{\rm Cl}}=3/2$, $I_H=1/2$). The selection rules $e\leftrightarrow e$, $f\leftrightarrow f$ for $\Delta J=\pm1$ and $e\leftrightarrow f$, $f\leftrightarrow e$ for $\Delta J=0$ thus give rise to $P$, $Q$ and $R$ branches with four lines for each rotational transition. These lines split into multiplets due to the hfs of the levels ($F=\left|J+I\right|...\left|J-I\right|$) arising from the interaction with the Cl nucleus spin.  In low-$J$ transitions, the hfs is clearly evident, although not completely resolvable, and its amplitude rapidly diminishes as $J$ increases. Figure \ref{fig:figf}a shows an example of lines where the hyperfine components cannot be resolved due to their proximity and the Doppler broadening of the lines. In turn, Figure \ref{fig:figf}b shows the $^2\Pi_{3/2}P(3/2)^f$ line of H$^{35}$Cl$^+$ line recorded at $\sim180$ K, where the hfs is almost completely resolved. There are intermediate cases where the hfs manifests as broadening or asymmetries of the lines, an example is shown in Figure \ref{fig:figf}c.  In this case, both $e$ and $f$ components of the $^2\Pi_{3/2}P(3/2)$ line of H$^{37}$Cl$^+$ overlap, and, together with the hfs, yield a complex, unresolved asymmetric profile.  The same line in the $\Omega=1/2$ ladder is split into $e$ and $f$ components, also with an asymmetric profile (the $f$ component is outside the frequency range of the figure).  It must be noted that the proton hfs could not be resolved in the THz study by \citet{Gupta12}.  In the present work, we access lines more sensitive to this interaction, but, as could be expected from the larger Doppler widths and the magnitude of the splittings predicted by theory \citep{Bru06}, we have not been able to resolve it either, and only the Cl hyperfine interaction has been considered.

From the line width of a Gaussian profile fitted to unresolved transitions with negligible hfs broadening, we have obtained kinetic temperatures of $\sim400$ K for the spectra recorded with water cooling, and $\sim270$ K for those with liquid nitrogen cooling.  The ratio of intensities of different $J$ lines under rotational equilibrium is also in agreement with this estimation.  The ratio of intensities of lines with the same $J$ in $\Omega=1/2,3/2$ leads to a much higher spin temperature, near $\sim750$ K, as can be inferred from the higher intensity of the $\Omega=1/2$ lines as compared with the calculations assuming Boltzmann equilibrium in Figures \ref{fig:figf}a and \ref{fig:figf}c.  This is not surprising since there is no reason to expect an equilibrium between rotational, vibrational, or spin temperatures in a plasma.  Moreover, \citet{Gupta12} also observed a high spin temperature, and the emission data from \citet{She72} showed vibrational excitation up to $v=9$.  No hot-band lines were recorded, thus we cannot give an estimate of the vibrational temperature of HCl$^+$ in the discharge.

The list of assignments and observed IR frequencies, together with those calculated and their residuals after the fit (see the Appendix), is given in Table \ref{Tab:linelist}.

\begin{deluxetable*}{cccccccclclcc}
\singlespace 
\tablecolumns{1}
\tablewidth{0pc} 
\tablecaption{Quantum number assignments, observed and calculated wavenumbers, and their differences for the $v=1-0$ band of HCl$^+$ in the ground electronic state.
\label{Tab:linelist} }
\tablehead{\colhead{$A({\rm Cl})^a$} & \colhead{$N'$} & \colhead{$J'+\onehalf$} & \colhead{$F'$} & \colhead{$N''$} & \colhead{$J''+\onehalf$} 
& \colhead{$F''$} & \colhead{Parity} & \colhead{$\nu_{\rm obs}$} & \colhead{$\sigma^b$}& & \colhead{$\nu_{\rm calc}$} &\colhead{$(\nu_{\rm obs}-\nu_{\rm calc})$}\\
\vspace{-1ex} & & & &  & & & & \colhead{/\wn} & \colhead{$/10^{-5}$ \wn} & & \colhead{/\wn}& \colhead{$/10^{-5}$ \wn} }
\startdata
  37 &  1&  1&  2& 2&  2&  2& $f-f$ & 2536.09722& 30  &   & 2536.09701& 21.3   \\
  37 &  1&  1&  2& 2&  2&  1& $f-f$ & 2536.10792& 36  &   & 2536.10792& -0.3   \\
  37 &  1&  1&  2& 2&  2&  3& $e-e$ & 2536.67608& 40  &   & 2536.67600& 7.9    \\
  35 &  1&  1&  2& 2&  2&  3& $e-e$ & 2538.49530& 30  &   & 2538.49519& 10.6   \\
  35 &  8&  9&   & 8&  9&   & $f-e$ & 2543.90798& 30  &   & 2543.90823& -25.3  \\
  35 &  8&  9&   & 8&  9&   & $e-f$ & 2544.52117& 30  &   & 2544.52074& 43.3   \\
 35 &  1&  1&  1& 1&  1&  1& $e-f$ & 2568.54858$^c$& 30  & * & 2568.54624& 33.9   \\
 35 &  1&  1&  1& 1&  1&  2& $e-f$ & 2568.54858$^c$& 30  & * & 2568.54893& -35.3  \\
\enddata
\tablecomments{ $N=J-\Omega$. The $e,f$ parity labeling follows the convention of \citet{bro75}. Only transitions with resolved hfs have $F$ quantum numbers assigned.\\
$^a$ Chlorine mass number. \\
$^b$ Experimental uncertainty (1$\sigma$). \\
$^c$ An asterisk(*) flags transitions fitted as intensity weighted doublets.\\
This table is published in its entirety in the electronic edition of the Journal.  A portion is shown here for guidance regarding its form and content.}  
\end{deluxetable*}

\subsection{Band constants}

A new combined isotope-independent fit with THz, optical, and IR data was carried out and is described in the Appendix.  Using the Dunham equations and the correlated error propagation formulae, the band parameters can be determined directly from the fit \citep{Yu_2012}, and are given in Table \ref{Tab:Band_Constants}. In this format, comparisons to other data analyses in the literature are possible. A few things are remarkable: (1) the value of $A_D0$ is in much better agreement with \citet{Bro79} than with any other study, including \citet{Gupta12}; (2) the values are generally of slightly lower precision than Gupta et al., opposite to the trend observed in the Dunham fit (see the Appendix.) The former result is presumably due to Brown's careful approximations of higher-order parameters. The latter may be due to truncation error in the prior analysis in which parameter uncertainties are underestimated when  higher-order parameters that are not well approximated as zero are not included in the analysis.  When these terms are introduced they have uncertainty and this correlates with the prior uncertainties.

\onecolumngrid
\newpage
\begin{deluxetable}{rrrrrr}
\tablecolumns{6}
\tablewidth{0pc} 
\tablecaption{Spectroscopic constants of H$^{35}$Cl$^+$ in the $X^{2}\Pi$ state (in MHz). \label{Tab:Band_Constants}} 
\tablehead{ \colhead{Constant} & \colhead{This work} & \colhead{\citet{Gupta12}} & \colhead{\citet{Bro79}} & \colhead{\citet{Sae76}} & \colhead{\citet{Bru06}}}
\startdata 
$\nu_0$      & 77003804.2(21)    & 77003551(75)    &                  & 77005284(99)        &            \\
$\gamma_0$   & $-$8882(40)       & $-$9163(34)     & $-$8993(165)     &                     & $-$9114    \\
$\gamma_1$   & $-$8882(40)       & $-$9163(34)     &                  &                     &            \\
$A_0$        & $-$19279455.6(47) & $-$19279138(38) & $-$19278775(1000)& $-$19430419 (92)    & $-$18137444\\
$A_1$        & $-$19264103.2(47) & $-$19263430(69) &                  & $-$19417878(138)    &            \\
$A_{D0}$     & 45.0(12)          & 54.33(99)       & 45.3(42)         & 62.60(48)           &            \\
$A_{D1}$     & 68.9(12)          & 80.98(113)      &                  & 84.03(75)           &            \\
$A_{H0}$     & 0.01853(113)      &                 &                  &                     &            \\
$A_{H1}$     & 0.01853(113)      &                 &                  &                     &            \\
$B_0$        & 293443.752(64)    & 293444.013(48)  & 293420.5(60)     & 293607.7(23)        &            \\
$B_1$        & 283806.268(103)   & 283812.603(324) &                  & 283959.2(30)        &            \\
$D_0$        & 16.39011(163)     & 16.39759(157)   &                  & 16.419(13)          &            \\
$D_1$        & 16.16813(186)     & 16.20489(518)   &                  & 16.161(18)          &            \\
$10^4H_0$    & 4.823(103)        & 4.468(126)      &                  & 4.57(23)            &            \\
$10^4H_1$    & 4.823(103)        & 4.468(126)      &                  & 4.22(32)            &            \\
$p_0$        & 18281.2(10)       & 18280.17(104)   & 18279.60(94)     & 18259(12)           &            \\
$p_1$        & 17983.9(11)       & 17972.6(471)    &                  & 17953(17)           & 17598      \\
$p_{D0}$     & -1.266(24)        & $-$1.290(25)$^b$&                  & $-$0.975(51)        &            \\
$p_{D1}$     & -1.266(24)        & $-$1.290(25)    &                  & $-$0.932(98)        &            \\
$10^4p_{H0}$ & 5.80(150)         & 6.16(172)       &                  &                     &            \\ 
$10^4p_{H1}$ & 5.80(150)         & 6.16(172)       &                  &                     &            \\
$q_0$        & $-$332.76(12)     & $-$332.198(87)  & $-$343.0(11)     & $-$335.1(18)        & $-$360     \\
$q_1$        & $-$320.03(18)     & $-$320.72(210)  &                  & $-$318.7(33)        &            \\
$q_{D0}$     & 0.04052(326)      & 0.04106(478)    &                  & 0.0470(60)          &            \\
$q_{D1}$     & 0.04052(326)      & 0.04106(478)    &                  & 0.017(12)           &            \\
$h^a$        & 357.2(18)         & 357.04(60)      &                  &                     & 369.5      \\
$a$          & 409.10(113)       & 409.16(88)      &                  &                     & 408        \\
$b^a$        & 167.4(38)         & 169.6(34)       &                  &                     & 166        \\
$b_F$        & 77.0( 36)         & 78.34(254)      &                  &                     & 85         \\
$c$          & -271.2( 51)       & -274.64(313)    &                  &                     & -243       \\
$d$          & 521.56(256)       & 523.12(282)     &                  &                     & 489        \\
$10^3d_D$    & 88.8(540)         & 60.6(414)       &                  &                     &            \\
$eQq_0$      & -9.67(301)        & -8.03(210)      &                  &                     & -12.5      \\
$eQq_2$      & -351.6(272)       & -370.4(184)     &                  &                     & -169       \\
\enddata
\tablecomments{Due to lack of expansion, many higher order terms, in both $v=0$ and $v=1$,  are equivalent to their corresponding Dunham term ($\gamma=Y^{\gamma}_{00}$, $H=Y_{03}$, $A_{H}=Y^A_{02}$, $p_D=Y^p_{01}/2$, $p_H=Y^p_{02}/2$ and $q_D=Y^q_{01}/2$.)  Hyperfine structure parameters are the same for $v=0$ and $v=1$.} \tablenotetext{a}{Probably a misprint in \citet{Gupta12}.\tablenotetext{b}{Calculated using $h=a+(b+c)/2$ and $b=b_F-c/3$.}}
\end{deluxetable}

\newpage
\twocolumngrid

\section{Opacities} 
Using the Einstein $A$ coefficient for the  $v=0\leftarrow 1$ transition of 205.6 s$^{-1}$ calculated by \cite{Pra91}, we can obtain the vibrational transition dipole moment (0.1959 D) and the individual line intensities as a function of temperature.  With this information, it is possible to estimate the detectable column densities in hypothetical IR observations from the ground. For 5 K  (the estimated excitation temperature of HCl$^+$ in the clouds toward W31C and W94N  \citep{DeLuca2012}), the strongest IR absorption of $^{35}{\rm HCl}^+$ would be that of the $^2\Pi_{3/2}\ Q(3/2)$ transition at 2568.21258 \wn (76,993,076 MHz). Neglecting hfs and $\Lambda$ doubling  (that would not be observed for this line even with resolving powers R$\gg$100000), we obtain the following integrated intensities (in units of cm$^2$\,km\,s$^{-1}$molec$^{-1}$) $\alpha(T)=2.874\times10^{-15},\ 2.743\times10^{-15} \ {\rm and}\ 1.975\times10^{-15}$ for temperatures $T=5,\ 20\ {\rm and}\ 50$ K, respectively. \citet{DeLuca2012} estimated the column densities to be $N(^{35}{\rm HCl}^+)=8.2\pm2.5\times10^{13}$ cm$^2\,$molec$^{-1}$ for W31C and  $N(^{35}{\rm HCl}^+)=8.7\pm2.6\times10^{13}$  cm$^2\,$molec$^{-1}$ for W49N. The observed absorption peak intensity  will depend inversely on the observed linewidth, that, in turn, will depend on the velocity spread of the cloud and the spectrograph resolution. \citet{DeLuca2012} found $\Delta v\sim 1.5-4.5$ km\,s$^{-1}$ from the analysis of the hfs of the lines of HCl$^+$, in good agreement with the $\Delta v=4.2\pm1.5$ km\,s$^{-1}$ derived by \citet{God12} from SH$^+$ and CH$^+$ in the same and other similar diffuse clouds. For an instrument line width of 3 km\,s$^{-1}$ (that of spectrographs with a resolving power of $R=100,000$, as available from e.g. CRIRES+ \footnote{expected to be fully operational in 2018} at the  VLT in Paranal or TEXES at Mauna Kea) the combined line width would be $\sim$ 6 km\,s$^{-1}$, and one can expect peak absorptions of 3.85, 3.60, and 2.64 \%, respectively, for the above temperatures and a column density of $N=8\times10^{13}$  cm$^2$\,molec$^{-1}$. Infrared emission from the dust envelope around the proto-stars in those sources can be bright enough in the IR, and, indeed, the absorption of neutral HCl toward CRL2136 at $3-4\ \mu m$ with peak intensities of $\sim2\%$ has been reported by \citet{Goto13}, using the CRIRES instrument  at the VLT. Therefore, the absorptions estimated above should be clearly detectable against bright IR sources in a reasonable amount of observing time. \\

\section{Concluding remarks}
We have provided accurate wavenumber measurements of 183 vibration-rotation lines of H$^{35}$Cl$^+$ and H$^{37}$Cl$^+$, measured with a difference-frequency laser spectrometer in a hollow-cathode discharge cell. The new wavenumbers have improved the previous Dunham-type fit to optical and millimiter-wave data, allowing more accurate predictions of other transitions. Notably, the wavenumbers of the $^2\Pi_{3/2}\ Q(3/2)$) transitions of the $v=1-0$ band are  2568.21258$\pm0.00031$ and   2566.34959$\pm0.00032$ \wn ($\pm 3\sigma$) for H$^{35}$Cl$^+$ and H$^{37}$Cl$^+$, respectively. These wavenumbers should help in future searches for these molecules.  In the conditions of the diffuse clouds where this ion has been observed with {\it Herschel}, absorptions of a few percent should be observable from the ground with high-resolution IR telescopes.

\acknowledgements JLD, JC, VJH and IT acknowledge the financial support from the Spanish MINECO through the Consolider ASTROMOL project, grant CSD2009-00038, and from the ERC through the Synergy Grant ERC-2013-SyG NANOCOSMOS. Additional support through grants FIS2012-38175 (JLD), FIS2013-408087-C2-1P (VJH, IT) and AYA2012-32032 (JC) is also acknowledged. We thank J.R. Goicoechea for his reading and comments on the manuscript. Portions of the research described in this paper were performed at the Jet Propulsion Laboratory, California Institute of Technology, under contract with the National Aeronautics and Space Administration. Government sponsorship acknowledged.

\appendix
\section{Combined isotope-independent fit}
\floattable
\begin{deluxetable}{rrr}
\tablecolumns{3}
\tablecaption{Dunham parameters of HCl$^+$ in the $X^{2}\Pi$ state (in MHz.  Uncertainties are expressed in units of the last significant digit). \label{Tab:Dunham_Parameters}}
\tablehead{ \colhead{Parameters} & \colhead{Optical+THz+IR} & \colhead{Optical+THz}\\&\colhead{This work}& \colhead{\citep{Gupta12}}} 
\startdata 
 $\mu^{-1/2}U_{10}/2$     &  40117164(205)& 40117378(217)  \\
 $\delta_{10}^{\rm H}/2$  &    -10997( 54)& -11111( 82)    \\
 $\delta_{10}^{\rm Cl}/2$ &     250.2(131)&                \\
 $Y_{20}/4$               &   -405371( 86)& -405382( 90)   \\
 $Y_{30}/8$               &    1305.3( 99)& 1310.5(107)    \\
 $\mu^{-1}U_{01}$         &  298527.4( 47)& 298529.7( 47)  \\
 $\mu^{-3/2}U_{11}/2$     &  -4902.66(100)& -4898.54(122)  \\
 $\delta_{01}^{B,\rm H}$  &    -210.3( 47)& -216.6( 48)    \\
 $Y_{21}/4$               &    18.202(197)& 17.797(240)    \\
 $\delta_{11}^{B,\rm H}/2$&     11.11( 63)& 11.65( 91)     \\
 $\mu^{-2}U_{02}$         &  -16.5654(100)& -16.5593(100)  \\
 $Y_{12}/2$               &   0.11099( 37)& 0.09637(248)   \\
 $\delta_{02}^{D,\rm H}$  &    0.0643( 97)& 0.0653(101)    \\
 $10^{3}Y_{03}$           &    0.4823(103)& 0.4468(126)    \\
 $Y^{\gamma}_{00}$        &     -8882( 40)& -9163(34)      \\
 $Y_{00}^A$($A_e$)        & -19289074(211)& -19288744(222) \\
 $Y^A_{10}/2$             &     11468(341)& 10596(377)     \\
 $\delta_{10}^{A,\rm H}/2$&      -691( 58)&                \\
 $Y^A_{20}/4$             &     -1329(140)&-1124(160)      \\
 $Y^A_{30}/8$             &     170.4(164)& 134.9(195)     \\
 $Y^A_{01}$               &     33.33(120)& 41.02(114)     \\
 $Y^A_{11}/2$             &   11.8413(298)& 13.32(39 )     \\
 $Y^A_{02}$               &   0.01852(113)&                \\ 
 $Y^p_{00}$               &  -9214.94( 53)& -9217.0(33)    \\
 $Y^p_{10}/2$             &    74.331(163)& 76.9( 32)      \\
 $Y^p_{01}$               &    0.6328(118)& 0.6448(127)    \\
 $10^{3}Y^p_{02}$         &    -0.290( 74)& -0.308( 86)    \\
 $Y^q_{00}/2$             &  -169.566( 74)& -168.97( 54)   \\
 $Y^q_{10}/2$             &     3.184( 36)& 2.87( 53)      \\
 $Y^q_{01}$               &   0.02026(163)& 0.02053(239)   \\
\enddata 
\end{deluxetable}

A full isotope-independent analysis including the fine-structure and magnetic hfs Hamiltonian was carried out utilizing the SPFIT program \citep{spfit_1991}. The full Hamiltonian and its mass and magnetic relationships among parameters are described in the Appendix of \citet{Gupta12}, in \citet{Dro01} and in \cite{Bro79}.  Note that in \citet{Gupta12} the signs should be reversed in the definition of $X_v$ parameters involving $p$ and $q$ related terms, and that the two rightmost factors in Equations A4 and A5 should be removed. The new fit increases the number of fitted lines from 8137 to 8324 (8103 lines from the optical data from \citet{She72}; 34 lines from \citet{Gupta12} and 187 lines from this work). 
The present data were weighted according to their estimated uncertainty.  Lines that showed no significant broadening or asymmetries were fitted to Gaussian profiles to obtain the line center, while for clearly asymmetric lines the centroid frequency was used. Some blends were appropriately intensity weighted and severely blended lines, where $e/f$ components could not be resolved, were not used.  The new assignments, observed and calculated frequencies, and their residuals are shown in Table \ref{Tab:linelist}.

The present high-precision data, together with the high-precision THz data of \citet{Gupta12}, provide anchor points for an improved analysis of the molecular parameters for the $^2\Pi$ state, which are shown in Table \ref{Tab:Dunham_Parameters} as determined by \citet{Gupta12} and presently. The precision of many of the Dunham parameters is improved, especially the higher-order $p,q$ terms that were previously completely dependent on the optical data. Unlike some of the more heavily correlated parameters associated with the spin-orbit distortion and electron-spin rotation, the $p$ and $q$ terms from the prior analysis do not deviate much outside the prior uncertainty (all less than 1.1$\sigma$). The precision afforded by the new data refines these values by factors of 6.1 and 7.3 for $Y^p_{00}$ and $Y^q_{00}$, respectively, and by factors of 2.0 and 1.5 for $Y^p_{10}$ and $Y^q_{10}$, respectively. The most dramatically modified parameters in the present fit are, not surprisingly, the most heavily correlated ones, $Y^{\gamma}_{00}$ and $Y^A_{01}$, which change by about 7-8$\sigma$ outside the previously determined error bars. The changes stem from the new sensitivity to the $Y^A_{11}/2$ parameter, which changes by about 4$\sigma$ outside its prior uncertainty and is determined 13$\times$ more precisely. The combined changes in $Y^A_{01}$ and $Y^A_{11}/2$ produce the anti-correlated change in $\gamma$, which is then further removed from its near equivalence with $Y^p_{00}$ and thus implies that HCl$^+$ is less consistent with the pure-precession hypothesis, but only by 3\%.
The basic Dunham parameters and the associated BOB terms are fairly stable upon addition of the higher fidelity vibrational data. This is seen in $\delta_{10}^H/2$ and $\delta_{01}^H$, which can be very sensitive to changes in the vibrational data set, but here only change by $<$1.5$\sigma$. This indicates that the previous data set achieved a broad enough scope in vibration and mass to accurately determine these subtle effects. However, higher-order terms in the Dunham model remain susceptible to the breadth and scope of the data and statistically significant changes occur in many of these terms. The only other parameters to have changed by more that 3$\sigma$ of its prior uncertainty are $\mu^{-3/2}U_{11}/2$ and $Y_{12}/2$, which appear to have correlated (and anti-correlated) commensurate changes with $\delta_{01}^H$, which is not outside of 3$\sigma$ of the prior uncertainty due to the larger uncertainty in that parameter. Due to correlation, any improvement in the precision of these two parameters is not obvious.  New frequency predictions for all isotopologues are available at 
\href{http://spec.jpl.nasa.gov/ftp/pub/catalog/catdir.html}{http://spec.jpl.nasa.gov/ftp/pub/catalog/catdir.htm}.

\end{document}